\documentclass[twocolumn,superscriptaddress,showpacs,aps,a4paper,floatfix,10pt,pra,a4paper]{revtex4-1}
\usepackage{graphicx}
\usepackage{amssymb}
\usepackage{amsmath}
\usepackage{color}
\usepackage{psfrag}
\usepackage{epsfig}
\usepackage{bbm}
\usepackage{bm}
\usepackage{hyperref}
\hypersetup{breaklinks}

\newcommand{\rb}[1]{\left( #1 \right)}
\newcommand{\ew}[1]{\langle #1 \rangle}
\newcommand{\beq}{\begin{eqnarray}}
\newcommand{\eeq}{\end{eqnarray}}
\newcommand{\eq}[1]{Eq.\,(\ref{#1})}
\newcommand{\fig}[1]{Fig.\,\ref{#1}}
\newcommand{\secref}[1]{Sec.\,\ref{#1}}

\newcommand{\citer}[1]{{Ref.~\cite{#1}}}

\begin{document}

\title{Energy relaxation in hot electron quantum optics via acoustic and optical phonon emission}
\author{C.~Emary}
\author{L.~A.~Clark}
\affiliation{Joint Quantum Centre (JQC) Durham-Newcastle, School of Mathematics, Statistics and Physics, Newcastle University, Newcastle-upon-Tyne, NE1 7RU, United Kingdom}

\author{M.~Kataoka}
\affiliation{
 National Physical Laboratory, Hampton Road, Teddington, Middlesex TW11 0LW, United Kingdom
}

\author{N.~Johnson}
\affiliation{NTT Basic Research Laboratories, NTT Corporation, 3-1
Morinosato Wakamiya, Atsugi, Kanagawa 243-0198, Japan
}

%%%%%%%%%%%%%%%%%%%%%%%%%%%%%%%%%%%%%%%%%%%%%%%%%%%%%%%%%%%%%%%%%%%%%%%
\begin{abstract}
    We study theoretically the relaxation of hot quantum-Hall edge-channel electrons under the emission of both acoustic and optical phonons.  Aiming to model recent experiments with single-electron sources, we describe simulations that provide the distribution of electron energies and arrival times at a detector a fixed distance from the source.
    From these simulations we extract an effective rate of emission of optical phonons that contains contributions from both a direct emission process as well as one involving inter-edge-channel transitions that are driven by the sequential emission of first an acoustic- and then an optical- phonon.
    Furthermore, we consider the mean energy loss due to acoustic phonon emission and resultant broadening of the electron energy distribution and derive an effective drift-diffusion model for this process.
\end{abstract}

\maketitle
%%%%%%%%%%%%%%%%%%%%%%%%%%%%%%%%%%%%%%%%%%%%%%%%%%%%%%%%%%%%%%%%%%%%%%%

%%%%%%%%%%%%%%%%%%%%%%%%%%%%%%%%%%%%%%%%%%%%%%%%%%%%%%%%%%%%%%%%%%%%%%%%%
%%%%%%%%%%%%%%%%%%%%%%       introduction      %%%%%%%%%%%%%%%%%%%%%%%%%%
%%%%%%%%%%%%%%%%%%%%%%%%%%%%%%%%%%%%%%%%%%%%%%%%%%%%%%%%%%%%%%%%%%%%%%%%%
\section{Introduction}
Dynamical quantum dots have emerged as accurate, on-demand, sources of single
electrons \cite{Blumenthal2007, Kaestner2008, Leicht2011, Kataoka2011, Giblin2012, Fletcher2012, Fletcher2013, Waldie2015}. 
One particularly novel feature of these charge pumps, as compared with other single-electron sources \cite{Feve2007, Kataoka2009, McNeil2011, Hermelin2011, Dubois2013}, is the high energy at which they inject electrons into quantum Hall edge channels.  This opens up the possibility of performing quantum-optics-like experiments with electrons \cite{Ji2003,Litvin2007,Roulleau2007,Roulleau2008,Roulleau2008a, Roulleau2009,Bieri2009, Bocquillon2012, Ubbelohde2015, Freulon2015,Tewari2016} in a new energy regime.
Clearly, the success of such experiments will depend critically on the relaxation and decoherence properties of the hot electrons as they are transmitted along edge channels.
At low injection energies, electron-electron interactions will dominate the relaxation of hot electrons \cite{Taubert2011}.  However, at high energies and fields, the transport electrons become ever-more localised in the edge of the sample and their interaction with the cold, bulk electrons becomes much reduced.  This then opens up the possibility that other relaxation mechanisms, principally phonon emission, will become the limiting inelastic mechanism.

Longitudinal-optical (LO) phonon emission from the hot electrons emitted by dynamic-quantum-dot single-electron sources has been observed in \citer{Kataoka2011, Giblin2012, Fletcher2012, Fletcher2013, Waldie2015}, and most recently studied in detail in \citer{Johnson2018}.  The theory of direct LO phonon emission within a Fermi's golden rule approach was discussed in \citer{Emary2016}, the key prediction of which was that increasing magnetic field strength should dramatically suppress LO-phonon emission.
In \citer{Johnson2018}, however, rates of LO-phonon emission extracted from time-of-flight and survival-probability measurements were seen to be significantly greater than those predicted by this theory.  The mechanism proposed to explain this was that the observed LO emission was in fact a sequential two-phonon process, where the electron first emits a longitudinal acoustic (LA) phonon and then the LO phonon.

In this paper we investigate more fully the role of acoustic phonons in the relaxation of hot single electrons, and focus on LA phonons interacting via the deformation potential interaction, which is expected to be dominant in GaAs heterostructures \cite{Bockelmann1990}.   We consider both LA and LO phonons and calculate the respective rates for emission by hot electrons in quantum Hall edge channels.  These rates then form the basis for the simulations of single electrons injected as localised wave packets into a quantum Hall edge channel.
These simulations allow us to calculate the energy-resolved arrival-time distribution (ATD) of the electrons at a detector some fixed distance from the electron source \cite{Fletcher2013, Waldie2015}.  
From this we extract survival probabilities, mean time-of-flight and also the effective LO emission rate as in \citer{Johnson2018}.  We study the energy and field dependence of this effective rate, and highlight contributions from direct LO emission processes and the sequential LA+LO channel.  We show how, in the regime dominated by the sequential channel, the effective LO rate is given by the total LA emission rate for inter-edge-channel scattering \cite{Komiyama1992}.

We then consider the impact of LA-phonon emission in its own right, and in particular its influence on the distribution of electron energies observed at the detector.  We show how this mechanism leads to a drop in mean energy of the electrons, as well as an increase in the width of their distribution in energy.  Moreover, we show that these features can be captured by an analytically-solvable  drift-diffusion model and derive explicit expressions for behaviour of the energy cumulants.  One key prediction of this model is that, at high energies, the width of electron energy distribution should be proportional to the square of the mean time-of-flight of the electrons.  Furthermore, we show that the effects of acoustic phonons on the electron energy distribution increases algebraically with magnetic field.

With our focus on single-electron sources and observables pertinent to current experiments, our work differs considerably from earlier works on phonon emission in quantum Hall systems, e.g. Refs.~\cite{Komiyama1992,Xu1996}.

This paper is structured as follows. In \secref{SEC:EPint} we set up our model of  quantum-Hall electrons interacting with LO and LA phonons. In 
\secref{SEC:rate} we derive from this model phonon emission rates and write down a master equation.  In \secref{SEC:ATD} we describe our simulations and the calculation of electron arrival-time distributions.  \secref{SEC:LO} contains our discussion of the effective LO-phonon emission rate, and \secref{SEC:DD} details our drift-diffusion model for relaxation under acoustic phonon emission.  We finish with discussions in \secref{SEC:discussions}.  A number of calculational details are given in the appendix.

%%%%%%%%%%%%%%%%%%%%%%%%%%%%%%%%%%%%%%%%%%%%%%%%%%%%%%%%%%%%%%%%%%%%%%%%%
%%%%%%%%%%%%%%%%%%%%%%%%%%%%%%%%%%%%%%%%%%%%%%%%%%%%%%%%%%%%%%%%%%%%%%%%%
%%%%%%%%%%%%%%%%%%%%%%%%%%%%%%%%%%%%%%%%%%%%%%%%%%%%%%%%%%%%%%%%%%%%%%%%%
\section{Electron-phonon interactions \label{SEC:EPint}}

We consider a Hamiltonian $H = H_\mathrm{e} + H_\mathrm{p} + V_\mathrm{ep}$, where the three contributions respectively describe the electrons, the phonons and electron-phonon interactions.
The electrons we model as quasi-two-dimensional with strong confinement in the $z$-direction and weak harmonic confinement transverse to the transport direction.  In the presence of perpendicular magnetic field, strength $B$, electron states are described by the edge-channel index $n=0,1,2,\ldots $, and wave number $k>0$ in the transport direction \cite{Datta1997, Emary2016}.  The electron Hamiltonian reads
\beq
  H_\mathrm{e} = \sum_{nk} E_{nk} c^\dag_{nk} c_{nk},
\eeq
with energies measured from the bottom of the $n=0$ subband
\beq
  E_{nk} &=& 
  n \hbar\Omega
    +  \frac{1}{2} \rb{\frac{\Omega}{\omega_c}}^2 m_e^* \omega_y^2 y^2_\text{G}(k)
  \label{EQ:eenergies}
  .
\eeq
Here, $m_e^*$ is the effective electron mass,
$\omega_y$ is the parabolic confinement frequency,
$
  \Omega = \rb{\omega_y^2 + \omega_c^2}^{1/2}
$, $\omega_c = {|eB|}/{m_e^*}$ is the cyclotron frequency, and 
\beq
  y_\text{G}(k) =  \rb{\frac{\omega_c}{\Omega}}^2\frac{\hbar {k} }{eB} 
  ,
\eeq
is the ``guide centre'' of the transverse wave function.  The transverse extent of these wave functions is given by the length 
$
l_\Omega = \rb{\hbar/m_e^* \Omega}^{1/2}$, and the velocity of an electron with energy $E$ in edge-channel $n$ is
$ 
  v_n(E) 
  = 
  \rb{\omega_y/\Omega}\left[2\rb{E-n\hbar \Omega}/m_e^*\right]^{1/2}
$ \cite{Kataoka2016}.

We assume that the LO phonons are dispersionless with energy $\hbar \omega_{\mbox{\tiny LO}} =  36$\,meV \cite{Taubert2011} and that the LA phonons have linear dispersion with speed of sound $c_{\mbox{\tiny LA}}$. With annihilation operators $a_\mathbf{q}$ and $b_\mathbf{q}$ for LO and LA phonons of wave vector $\mathbf{q}$, the phonon Hamiltonian reads
\beq
  H_\mathrm{p} 
  =
  \hbar \omega_{\mbox{\tiny LO}} 
  \sum_{\mathbf{q}} 
  a_\mathbf{q}^\dag
  a_\mathbf{q}
  + 
  \hbar c_{\mbox{\tiny LA}}
  \sum_{\mathbf{q}}
  q \,
  b_\mathbf{q}^\dag
  b_\mathbf{q}
  .
\eeq
Specified in terms of $\tilde{c}_\mathbf{k}$, the annihilation operator for plane-wave electrons with three-dimensional wave vector $\mathbf{k}$, the Fr\"{o}hlich Hamiltonian for electron-phonon interactions \cite{Froehlich1954, Mahan2000} reads 
\beq
  V_\text{ep} &=& 
  \sum_{\mathbf{k},\mathbf{q}} 
  \mathcal{M}_{\mbox{\tiny LO}}(\mathbf{q})\,
  \tilde{c}_{\mathbf{k}+\mathbf{q}}^\dag \tilde{c}_\mathbf{k}
  \rb{
    a_{-\mathbf{q}}^\dag  
    + a_{\mathbf{q}}
  }
  \nonumber\\
  &&
  +
  \sum_{\mathbf{k},\mathbf{q}} 
  \mathcal{M}_{\mbox{\tiny LA}}(\mathbf{q})\,
  \tilde{c}_{\mathbf{k}+\mathbf{q}}^\dag \tilde{c}_\mathbf{k}
  \rb{
    b_{-\mathbf{q}}^\dag  
    + b_{\mathbf{q}}
  }
  \label{EQ:FH1}
  .
\eeq
The momentum dependence of the matrix elements is given by \cite{Mahan2000, Climente2006}
\beq
  \left|\mathcal{M}_{\mbox{\tiny LO}}(\mathbf{q}) \right|^2
  \equiv  
  \frac{M_{\mbox{\tiny LO}}^2}{L^3} 
  \frac{1}{q^2}
  ;~~
  \left|\mathcal{M}_{\mbox{\tiny LA}}(\mathbf{q})\right|^2 = \frac{M_{\mbox{\tiny LA}}^2}{L^3} q
  .
\eeq
where $q=|\mathbf{q}|$ and $L^3$ is the sample volume.  Details of the couplings $M_{\mbox{\tiny LO}}$ and $M_{\mbox{\tiny LA}}$ are given in Appendix~\ref{SEC:APmatrixeles}.  We have assumed here that the interaction with LA phonons is exclusively through the deformation-potential interaction (LADP for short) \cite{Gantmakher1987} [we have also considered the effects of piezo-electric phonon interactions but these are generally unimportant here  (see Appendix \ref{SEC:APPZ})].

In terms of the edge-channel states, the interaction reads
\beq
  V_\text{ep} &=& 
  \sum_{n n'}  \sum_{k k'}
  \sum_{\mathbf{q}}
    \Lambda_{n'k'nk}^{\mbox{\tiny LO}}(\mathbf{q})
    c_{n'k'}^\dag c_{nk}
    \rb{
    a_{-\mathbf{q}}^\dag  
    + a_{\mathbf{q}}
  }
  \nonumber\\
  &&
  \!\!+
  \sum_{n n'}  \sum_{k k'}
  \sum_{\mathbf{q}}
    \Lambda_{n'k'nk}^{\mbox{\tiny LA}}(\mathbf{q})
    c_{n'k'}^\dag c_{nk}
    \rb{
    b_{-\mathbf{q}}^\dag  
    + b_{\mathbf{q}}
  }  
  \label{EQ:V}
  ,
\eeq
with matrix elements $\Lambda_{n'k'nk}^{\mbox{\tiny LO}}$ and $\Lambda_{n'k'nk}^{\mbox{\tiny LA}}$ obtained as in \citer{Emary2016}.

%%%%%%%%%%%%%%%%%%%%%%%%%%%%%%%%%%%%%%%%%%%%%%%%%%%%%%%%%%%%%%%%%%%%%%%
%%%%%%%%%%%%%%%%%%%%%%%%%%%%%%%%%%%%%%%%%%%%%%%%%%%%%%%%%%%%%%%%%%%%%%%
%%%%%%%%%%%%%%%%%%%%%%%%%%%%%%%%%%%%%%%%%%%%%%%%%%%%%%%%%%%%%%%%%%%%%%%
\section{Relaxation rates and master equation \label{SEC:rate}}

We consider single electrons injected into the system, and since experiments are performed at low temperature, we consider phonon emission only.
Let $P_{nk}$ be the probability to find an electron in edge channel $n$ with wave number $k$.  Within a master equation approach \cite{Breuer2007}, the equation for the evolution of these probabilities reads
\beq
  \dot{P}_{nk} &=&
  -\sum_{n'k'\nu } 
    \Gamma^{\nu}_{n'k'nk} 
  P_{nk} 
  + \sum_{n'k'\nu} 
    \Gamma^{\nu}_{nkn'k'} 
  P_{n'k'}
  \label{EQ:ME}
  ,
\eeq
where $\Gamma^\nu_{n'k'nk} $ is the transition rate from state $nk$ to $n'k'$ induced by the emission of a $\nu = \mathrm{LO},\mathrm{LA}$ phonon.  According to Fermi's golden rule \cite{Gantmakher1987,Ridley2009}, these rates read
\beq
  \Gamma^{\nu}_{n'k'nk} 
  &=&
  \frac{2\pi}{\hbar}
  \sum_\mathbf{q}
  \left|\Lambda^{\nu}_{n'k'nk}\right|^2
  \delta\rb{
    E_{n'k'}-E_{nk} + \hbar \omega_{\nu}
  }
  \label{EQ:FGR1}
\eeq
with $\omega_\nu = \omega_{\mbox{\tiny LO}}$ for $\nu = \mathrm{LO}$ and
$\omega_\nu = \omega_{\mbox{\tiny LA}} = c_{\mbox{\tiny LA}} q$ for $\nu = \mathrm{LA}$.  The explicit evaluation of these quantities for the LADP interaction is discussed in Appendix~\ref{SEC:APrates}.  Details of the evaluation of the LO emission rates are as in \citer{Emary2016}.

Taking the continuum limit and defining the continuous distributions
$\rho_n(E)= \sum_k \delta(E - E_{nk}) P_{nk}$ in terms of electron energy $E$, our master equation becomes
\beq
  \dot{\rho}_n(E) 
  &=& 
  - \sum_{n'\nu}\int dE'\,
    \widetilde{\Gamma}^\nu_{n'n}(E',E)
   \rho_n(E) 
  \nonumber\\
  &&
  + \sum_{n'\nu}\int dE'\,
    \widetilde{\Gamma}^\nu_{nn'}(E,E')
  \rho_{n'}(E')
  \label{EQ:MEdE}
  ,
\eeq
with ``rate densities''
\beq
  \widetilde{\Gamma}^{\nu}_{n'n}(E',E) &\equiv &
  \frac{L}{2\pi} \frac{1}{\hbar v_0(E')}
  \int dE_k \int dE_{k'} 
  \nonumber\\
  &&  
  \times
  \delta\rb{E-E_k}\delta\rb{E'-E_{k'}}
  \Gamma^{\nu}_{n'k'nk}
  \label{EQ:ratedensity}
  .
\eeq
The total scattering rate out of state in subband $n$ with energy $E$ into subband $n'$ is given by the integral
\beq
  {\Gamma}^{\nu\, \mbox{\tiny tot}}_{n'n}(E)
  =
  \int dE' \widetilde{\Gamma}^{\nu}_{n'n}(E',E)
  .
\eeq

%%%%%%%%%%%%%%%%%%%%%%%%%%%%%%%%%%%%%%%%%%%%%%%%%%%%%%%%%%%%%%%%
\begin{figure}[t]
  \begin{center}
     \includegraphics[width=\columnwidth,clip=true]{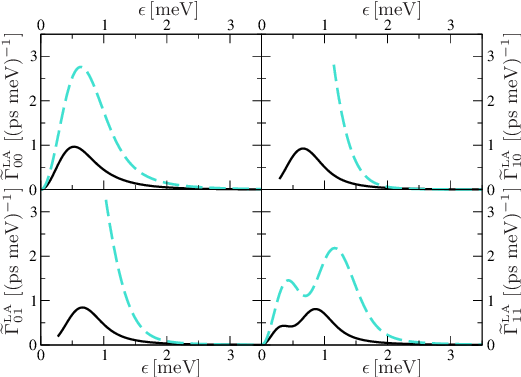}
  \end{center}
  \caption{
    The rate densities $\widetilde{\Gamma}^{\mbox{\tiny LA}}_{n'n}(E-\epsilon,E)$ as a function of energy loss $\epsilon$ for transitions induced by acoustic-phonon emission in the outermost two edge channels $n,n'=0,1$.  The initial energy $E$ was 100\,meV above the $n=0$ band bottom, and results are shown for two values of the magnetic field: $B=6$\,T (black solid), $B=12$\,T (blue dashed). Other parameters are as set out in Appendix~\ref{SEC:APmatrixeles}.  At these fields, the subband spacings are respectively $\hbar \Omega = 10.3$\,meV and $\hbar \Omega=20.9$\,meV.  The rate densities are defined such that the area under each curve gives the total rate out of a state with energy $E$ and subband $n$ into subband $n'$.  The curves for off-diagonal rates end abruptly at low $\epsilon$ due to the cut-off \eq{EQ:constraint}.
    \label{FIG:LArates}
  }
\end{figure}
%%%%%%%%%%%%%%%%%%%%%%%%%%%%%%%%%%%%%%%%%%%%%%%%%%%%%%%%%%%%%%%%

\fig{FIG:LArates} shows these rate densities for transitions in the outermost two edge channels with magnetic field strengths of $B=6$\,T and $B=12$\,T.
For all transitions, we observe that LADP phonon scattering is only significant for changes in electron energy $\epsilon = E-E'$ of a few meV.  In addition, the rate densities at $6$\,T are smaller than their $12$\,T counterparts.  Given the greater electron velocity at lower fields, most of this difference arises from the difference in density-of-states factor $1/\rb{\hbar v_0}$ in \eq{EQ:ratedensity}.
The $n\ne n'$ rates that describe inter-edge-channel scattering \cite{Komiyama1992} are finite only above a certain minimum value of $\epsilon$ given by 
\beq
 \epsilon  \ge  \hbar c_{\mbox{\tiny LA}} |k-k'|
  \label{EQ:constraint}
  .
\eeq
This lower cut-off arises because of energy and momentum conservation and the fact that, when changing subband, a small drop in electron energy necessarily has associated with it a change in electron wave number.

%%%%%%%%%%%%%%%%%%%%%%%%%%%%%%%%%%%%%%%%%%%%%%%%%%%%%%%%%%%%%%%%%%%%%%%
%%%%%%%%%%%%%%%%%%%%%%%%%%%%%%%%%%%%%%%%%%%%%%%%%%%%%%%%%%%%%%%%%%%%%%%
%%%%%%%%%%%%%%%%%%%%%%%%%%%%%%%%%%%%%%%%%%%%%%%%%%%%%%%%%%%%%%%%%%%%%%%
\section{Arrival-time distribution and simulations\label{SEC:ATD}}

The stochastic nature of phonon emission means that different electrons emit a different sequence of phonons as they travel.  This, coupled with the energy-dependence of the electron velocity, means that after travelling a distance $x_D$ from source to detector, the electrons will arrive at a range of times.  The distribution of such times is called the arrival-time distribution, and currently the most general such object accessible in experiment is the energy-resolved arrival-time distribution $A(E,\tau,x_D)$, defined such that $A(E,\tau,x_D) \Delta E \Delta \tau$ is the probability that an electron arrives at detector position $x_D$ with an energy between $E$ and $E+\Delta E$  between times $\tau$ and $\tau + \Delta \tau$ (in the limit $\Delta E$, $\Delta t \to 0$).
The calculation of the ATD within quantum mechanics involves some subtleties \cite{Allcock1969,Dumont1993,Delgado1999,Muga2000}. However, we shall here pursue a semi-classical description \cite{Kataoka2017}, in which these issues do not arise.

We assume that the electrons are emitted in Gaussian wavepackets \cite{Ryu2016} centred around an energy $E=E_0$ and time $t=0$, with energy and time widths of $\sigma_E =1$\,meV and $\sigma_\tau=5$\,ps, in line with recent experiments \cite{Fletcher2013,Waldie2015,Johnson2017a}. Furthermore, we assume that these wave packets remain coherent during their transmission to the detector and that their shape remains constant.  The absence of significant dispersion over the relevant timescales was discussed in \cite{Emary2016}.  
The maintenance of coherence under phonon emission can be justified by considering the Bloch-Redfield equations for the coherences, analogous to those for the populations in \eq{EQ:ME}.  This analysis shows that the rates for the transfer of coherences to be approximately the same as those for the transfer of populations [\eq{EQ:FGR1}] when the spread in $k$ vectors is $\lesssim l_\Omega^{-1}$.  This is the case for the conditions studied here.

In this picture then, phonon emission transfers travelling, fixed-shape wavepackets between different energies.  What remains is to track the motion of the centre of these wavepackets as they travel, and this we do using a Monte Carlo simulation of individual electron trajectories based on the master equation, \eq{EQ:MEdE}.

% To calculate $A(E,\tau,x_D)$ from  \eq{EQ:MEdE} we use  

We first discretise the electron energy, $E_i = i \Delta E$; $i=0,1,2,\ldots$.  
In discrete time step $\Delta t$, then, an electron in band $n$ with energy $E_i$ has a probability to emit a phonon and scatter into a new state with $E_j$ in band $n'$ given by 
\footnote{ 
  We found that integration over the small energy interval is preferable to the using the approximation 
  $T_{ji}^{n'n} \approx \Delta t\,\Delta E\,{\widetilde{\Gamma}}_{n'n}(E_j,E_i)$ since this avoids problems with the cut-offs in \eq{EQ:constraint}, as well as with the rapid variations in the rates that occur in their vicinity.
}
\beq
  T_{ji}^{n'n} \approx \Delta t 
  \int_{E_j - \Delta E/2}^{E_j + \Delta E/2} dE'\,
  \widetilde{\Gamma}_{n'n}(E',E_i)
  .
\eeq
During this time step, an electron will also propagate a distance $\Delta x = v_n (E_i) \Delta t$.  
Thus, our procedure is to iterate these two steps of probabilistic phonon emission and deterministic electron propagation until the distance travelled reaches $x_D$. This is then repeated many times and the convolution of this set of wave-packet centres in $E$-$t$ space with the Gaussian wavepacket of the individual electrons builds up a picture of the energy-resolved ATD.

%%%%%%%%%%%%%%%%%%%%%%%%%%%%%%%%%%%%%%%%%%%%%%%%%%%%%%%%%%%%%%%%
\begin{figure}[tb]
  \begin{center}
     \includegraphics[width=\columnwidth,clip=true]{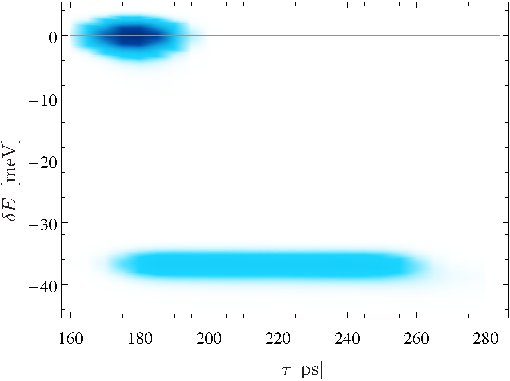}
  \end{center}
  \caption{
    Energy-resolved arrival-time distribution (ATD) $A(E,\tau,x_D)$ plotted as a function of energy loss $\delta E = E-E_{0}$, with $E_0$ the initial energy, and the arrival time $\tau$.  White indicates absence of electrons, darker colours greater electron probability density.
    Clustered around $\delta E = 0$ (horizontal line) is the ATD for electrons that arrive at the detector without having emitted any LO phonons.  Around $\delta E = -\hbar \omega_{\mbox{\tiny LO}} = -36$\,meV we see the ATD of the first LO-phonon replica.
    Parameters: $B=6$\,T, $E_0 = 70$\,meV, $x_D=28\,\mu\mathrm{m}$, $\sigma_E =1$\,meV and $\sigma_\tau=5$\,ps.  The number of electron trajectories followed was $10^5$.
    \label{FIG:flopLOW}
  }
\end{figure}
%%%%%%%%%%%%%%%%%%%%%%%%%%%%%%%%%%%%%%%%%%%%%%%%%%%%%%%%%%%%%%%%

For the results we present here, the electron always starts in the outermost edge-channel, $n=0$ \cite{Ryu2016}.  For numerical simplicity, our simulation only considers subbands $n=0,1$ where the majority of the dynamics takes place.  We consider the detector to be positioned a distance $x_D = 28\,\mu$m from the centre of the initial distribution.

\begin{figure}[tb]
  \begin{center}
     \includegraphics[width=\columnwidth,clip=true]{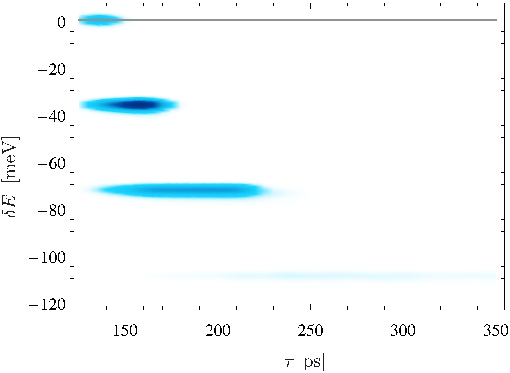}     
  \end{center}
  \caption{
    As \fig{FIG:flopLOW} but here  with the higher injection energy of $E_0 = 120$\,meV.
    Distributions corresponding to three LO-phonon replicas (the third is faint), plus that of the directly transmitted electrons, are observed.
    \label{FIG:flopHIGH}
  }
\end{figure}

\fig{FIG:flopLOW} and \fig{FIG:flopHIGH} show the energy-resolved ATD for two different starting energies $E_0=70,120$\,meV with a field of $B=6$\,T. Results are shown only for electrons detected in the $n=0$ edge channel.  This is by far the majority for these parameters since only $< 3\%$ of electrons end up in the $n=1$ band at the detector.
In both plots, a portion of the electron density remains clustered around the original injection energy $\delta E =  E-E_0= 0$ with central arrival time of $\tau = 176$\,ps for $E_0=70$\,meV (\fig{FIG:flopLOW}) and $\tau = 136$\,ps for $E_0 = 120$\,meV (\fig{FIG:flopHIGH}). These distributions represent those electrons that reach the detector without having emitted any LO phonons, and this part of the ATD we shall denote as the $A^{(0)}$ distribution.

The ATD also shows clusters positioned approximately around energy losses equal to multiples of the LO phonon energy.  These are the LO-phonon replicas, in which electrons have emitted $1,2,\ldots$ LO phonons en route to the detector.  In \fig{FIG:flopLOW} one replica is visible; in \fig{FIG:flopHIGH} there are three, and this number depends on both the initial energy as well as the distance of travel.  The initial energy sets the maximum number of phonon replica that can be observed since, once within $\hbar \omega_{LO}$ of the band bottom no further LO phonons can be emitted.  The distance travelled determines the degree to which each of these phonon replica are actually realised, with lower-energy replica becoming more populated the further the electron travels.
As is clear from these figures, the phonon-replica distributions arrive later than the original distribution due to the energy lost to LO phonons. They are also broadened along the time axis, which is due to the uncertainty in the emission time of the LO phonons and corresponding uncertainty in the fraction of total distance travelled in lower-energy states.

%%%%%%%%%%%%%%%%%%%%%%%%%%%%%%%%%%%%%%%%%%%%%%%%%%%%%%%%%%%%%%%%
%%%%%%%%%%%%%%%%%%%%%%%%%%%%%%%%%%%%%%%%%%%%%%%%%%%%%%%%%%%%%%%%

%%%%%%%%%%%%%%%%%%%%%%%%%%%%%%%%%%%%%%%%%%%%%%%
\section{Effective LO-phonon emission rate \label{SEC:LO}}

The LO-phonon rate analysis of \citer{Johnson2018} was based on estimation of the ``survival probability'', i.\,e. the probability of reaching the detector without having emitted a LO phonon or, in other words, the total weight of the $A^{(0)}$ distribution.
Although a counting-field approach would permit the calculation of exactly that part of the distribution having emitted $m$ LO phonons, we elect here instead to use a procedure based on inference from the ATD that matches the experimental procedure.
With only LO phonon processes active, it is a simple matter to infer this probability from the energy distribution of the electrons at the detector --- surviving electrons reach the detector with exactly the same energy with which they are injected.  
In the presence of the continuous energy loss from LA phonon emission, however, this picture is complicated by the fact that the electron energy distributions drop and broaden as they transit.
Therefore we here define ``survival'' to mean that the energy of the electron at the detector satisfies $E > E_\mathrm{surv} =E_0-\hbar \omega_{LO} + 10$\,meV.  The choice of $10$\,meV here is somewhat arbitrary, but the value should be large enough to ensure we avoid the tails of the emitted distribution and yet small enough that the $A^{(0)}$ distribution does not drop near this line under the emission of acoustic phonons. The value of $10$\,meV was found to perform well in both these respects. 
In terms of the ATD, the survival probabilities are obtained as
\beq
  P_n = 
  \int dE \,
  \int d \tau \,
  A_n(E,\tau,x_D)
  \Theta(E-E_\mathrm{surv})
  ,
\eeq
where $P_n$ and $A_n$ are quantities conditioned on finding the electrons in edge channel $n$.  Thus, $P_0$ is the survival probability with the electron being detected in the outermost edge channel; $P_1$ is the same but with the electron being detected in the $n=1$ channel.

%%%%%%%%%%%%%%%%%%%%%%%%%%%%%%%%%%%%%%%%%%%%%%%%%%%%%%%%%%%%%%%%
\begin{figure}[t]
  \begin{center}
     \includegraphics[width=\columnwidth,clip=true]{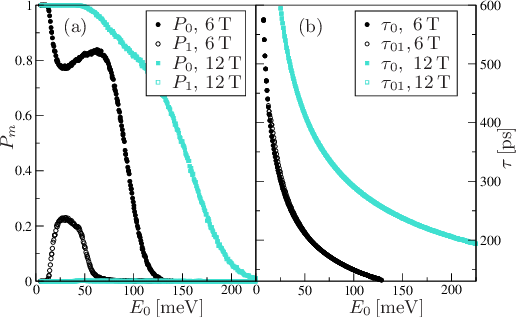}
  \end{center}
  \caption{
    (a) The survival probabilities $P_0$ and $P_1$ as a function of injection energy $E_0$ for two magnetic field strengths, $B=6,12$\,T. 
    (b) Mean time-of-flight for electrons in the $A^{(0)}$ distribution, i.\,e. those that reach the detector without having emitted a LO phonon.  $\ew{\tau}_0$ counts only those electrons detected in the outermost edge channel, whereas $\ew{\tau}_{01}$ counts those in both $n=0,1$ states. The difference between these two quantities is seen to be minor.
    \label{FIG:SP}
  }
\end{figure}
%%%%%%%%%%%%%%%%%%%%%%%%%%%%%%%%%%%%%%%%%%%%%%%%%%%%%%%%%%%%%%%%

\fig{FIG:SP}(a) shows survival probabilities $P_{0,1}$ as a function of injection energy for $B=6,12$\,T.  At low energies, $P_0 \approx 1$, with the vast majority of electrons reaching the detector without having emitted an LO phonon and being detected in the original edge channel.  As injection energy increases above $E_0>\hbar \omega_{\mbox{\tiny LO}}$, however, $P_0$ drops rapidly as LO emission becomes active.
The other striking feature in \fig{FIG:SP}(a) is that, at  low field, we see that there exists an energy range for which $P_1$ is significant (up to about 22\% in the figure). For higher fields, as exemplified by $B=12$\,T, however, this feature is almost completely absent.

If LO-phonon emission was the only process operative here then the survival probability $P_0$ would show an exponential decay as a function of time with rate given by the LO-phonon emission rate \cite{Emary2016}.  Let us therefore assume that a similar relation exists in the presence of LA-phonon emission and write an exponential relation between the survival probability,
$
  P_0 \approx e^{-\gamma_0 t}
$,
where $\gamma_0$ is an effective rate parameter describing the total decay of the survival probability.   We then approximate the time $t$ in this expression with $\ew{\tau}_{0}$ the mean time-of-flight for electrons in the $A^{(0)}$ distribution and outermost edge channel (this is shown in \fig{FIG:SP}(b)).  Thus, the expression we use to extract the effective LO emission rate is
\beq
  \gamma_0 = -\ew{\tau}_{0}^{-1} \log P_0
  \label{EQ:totalrate}
  .
\eeq
We also define $\gamma_{01} = -\ew{\tau}_{01}^{-1} \log (P_0+P_1)$, which takes into account ``survived'' electrons in both $n=0$ and $n=1$ subbands.

%%%%%%%%%%%%%%%%%%%%%%%%%%%%%%%%%%%%%%%%%%%%%%%%%%%%%%%%%%%%%%%%
\begin{figure}[t]
  \begin{center}
     \includegraphics[width=\columnwidth,clip=true]{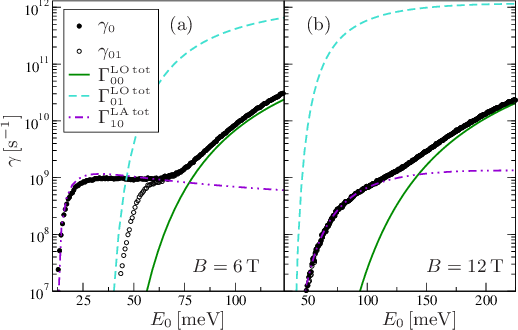}
  \end{center}
  \caption{
    The extracted LO-emission rates $\gamma_0$ (solid circles) and $\gamma_{01}$ (empty circles) as a function of injection energy for (a) $B=6$\,T, and (b) $B=12$\,T.  Also plotted are the calculated LO emission rates $\Gamma^{\mbox{\tiny LO\,tot}}_{\mbox{\tiny 00}}$, $\Gamma^{\mbox{\tiny LO\,tot}}_{\mbox{\tiny 01}}$ and total LA emission rate $\Gamma^{\mbox{\tiny LA\,tot}}_{\mbox{\tiny 10}}$.
    \label{FIG:gamma}
  }
\end{figure}
%%%%%%%%%%%%%%%%%%%%%%%%%%%%%%%%%%%%%%%%%%%%%%%%%%%%%%%%%%%%%%%%

Simulation results for $\gamma_0$ and $\gamma_{01}$ are shown in \fig{FIG:gamma} for both $B=6$\,T and $B=12$\,T.
At high energies, $\gamma_0$ and $\gamma_{01}$ are similar and both very well approximated by the bare LO phonon emission rate (green solid lines).  In this regime, therefore, we expect the conclusions of \citer{Emary2016} to hold.
At lower energies, however, the effective rates develop a ``knee'', which is more pronounced at lower fields and also more extreme in $\gamma_{0}$ than $\gamma_{01}$.  
Responsible for this enhancement is a relaxation process that involves first the emission of an LA phonon from the outermost edge channel to the $n=1$ channel, followed by the sequential emission of an LO phonon from $n=1$ back to the outermost edge channel. 
This process is best first analysed by looking at the $B=12$\,T case, \fig{FIG:gamma}b. Alongside simulation results ($\gamma_0$ and $\gamma_{01}$ are pretty much identical at this field), we plot the direct LO rates from $n=0$ to $n'=0$ (solid green lines) and from $n=1$ to $n'=0$ (dashed blue lines).  We also plot the total LA rate from $n=0$ to $n'=1$ (dot-dashed purple lines).
At high energy, $ \Gamma_{00}^{\mbox{\tiny LO\,tot}} \gg \Gamma_{10}^{\mbox{\tiny LA\,tot}} $ and the direct optical phonon process dominates.  Then, as the energy decreases there is a cross over in these rates and $\Gamma_{10}^{\mbox{\tiny LA\,tot}} \gg \Gamma_{00}^{\mbox{\tiny LO\,tot}}$.  Outscattering into the $n=1$ edge via LA emission then dominates over direct LO emission.  However, the rate $\Gamma_{01}^{\mbox{\tiny LO\,tot}}$ from $n=1$ back to $n=0$ is yet greater than all other rates considered and thus, upon arriving in the $n=1$ subband the electron rapidly emits an LO phonon and relaxes back into the $n=0$ subband.  The speed of this two-step process is governed by the slowest step, which is the LA phonon emission.  Thus, in the ``knee'' region, the effective LO rate $\gamma_0$ is well approximated by the total LA rate $\Gamma_{10}^{\mbox{\tiny LA\,tot}}$.  This approximation is observed to work well in \fig{FIG:gamma}b.
Furthermore, the rapidity of the $n=1$ to $n=0$ LO process also explains why for $B=12$\,T the population of the $n=1$ level at the detector is always small.

The story for $B=6$\,T is similar (\fig{FIG:gamma}a), with one significant difference.  The rate $\Gamma_{01}^{\mbox{\tiny LO\,tot}}$ drops rapidly and around an energy of $E_0 \approx 47$\,meV, falls below that of $\Gamma_{10}^{\mbox{\tiny LA\,tot}}$.  Below this point, there is no longer a rapid outscattering of electrons back from the $n=1$ to $n=0$ level.  This then leads to the finite occupancy of the $n=1$ level which, in \fig{FIG:SP}, is observed from $E_0 \approx 47$\,meV down until the rates drop off near the bandbottom.  This reduced outscattering also means that there is a significant difference between the effective rates $\gamma_0$ and $\gamma_{01}$ in this regime.

%%%%%%%%%%%%%%%%%%%%%%%%%%%%%%%%%%%%%%%%%%%%%%%%%%%%%%%%%%%%%%%%%%%%%%%%%%%%%%%%
%%%%%%%%%%%%%%%%%%%%%%%%%%%%%%%%%%%%%%%%%%%%%%%%%%%%%%%%%%%%%%%%%%%%%%%%%%%%%%%%
%%%%%%%%%%%%%%%%%%%%%%%%%%%%%%%%%%%%%%%%%%%%%%%%%%%%%%%%%%%%%%%%%%%%%%%%%%%%%%%%
\section{Acoustic-phonon-induced drift-diffusion in energy space \label{SEC:DD}}

Emission of LA phonons leads electrons to lose energy and, unlike LO emission, this occurs over a continuous range of energies.  Furthermore, the stochasticity of this process leads to a broadening of the electron distributions as they travel.  This broadening is just about apparent in \fig{FIG:flopLOW}, where the standard deviation of the $A^{(0)}$ distributions are greater than the starting value of 1\,meV; the shift in the centre of distribution is too slight to see from this figure for these parameters.

As in the last section, we here focus on the behaviour of the $A^{(0)}$ electrons and will write down an approximate analytic theory for the behaviour of this distribution.  To do so, we first discard LO processes (these we assume can be taken into account by an overall loss factor $e^{-\gamma_0 t}$ which, since $\gamma_0$ is effectively constant over the energy-scale relevant for LA emission, can simply be removed by normalising the $A^{(0)}$ distribution), and confine ourselves to the outermost $n=0$ edge channel.  To simplify notation, we thus drop the LA superscript and channel index for this section.

With injection energy much higher than the scale of any loss due to LA phonon emission, the dispersion relation can be linearised
\beq
  E_{nk} \approx \varepsilon^{(0)}_{n} + \hbar v_0 k
  \label{EQ:linearise}
  ,
\eeq
where $v_0$ is the electron velocity, here assumed constant and $\varepsilon^{(0)}_{n}$ is the energy offset. As shown in Appendix\,\ref{SEC:APdiff}, this linearisation means that the LA-phonon rate becomes dependent only on the energy difference and we thus write $\widetilde{\Gamma}(E',E) \to \widetilde{\Gamma}(E-E')$ for $E>E'$.  The relevant part of our master equation, \eq{EQ:MEdE}, then becomes
\beq
  \dot{\rho}(E) = 
  \int_0^{\infty} d\epsilon \, 
  \widetilde{\Gamma}(\epsilon) \left[ - \rho(E) + \rho(E+\epsilon)\right]
  \label{EQ:MEfinal}
  ,
\eeq
where we have extended the limits of integration commensurate with the linearisation.

\eq{EQ:MEfinal} can be solved by introducing the generating function of energy moments
\beq
  \rho(\chi,t) \equiv \int_{-\infty}^\infty dE\, e^{i\chi E} \rho\rb{E,t}
  .
\eeq
We then obtain
\beq
  \dot{\rho}(\chi,t) = 
  \left[\Lambda(\chi) - \Lambda(0)\right]\rho(\chi,t)
  \label{EQ:MEchi}
  ,
\eeq
with
\beq
  \Lambda(\chi)
  \equiv \int_0^\infty d\epsilon \, \widetilde{\Gamma}(\epsilon) e^{-i\chi \epsilon}
  .
\eeq
\eq{EQ:MEchi} can be solved to give 
\beq
   \rho(\chi,t) = e^{\left[\Lambda(\chi) - \Lambda(0)\right]t} \rho(\chi,0)
   ,
\eeq
such that the cumulant generating function for the system is
\beq
  \mathcal{F}(\chi,t) &\equiv& \ln  \rho(\chi,t) 
  = \left[\Lambda(\chi) - \Lambda(0)\right]t + \mathcal{F}(\chi,0)
  ,
\eeq
where $\mathcal{F}(\chi,0)$ describes the energy-statistics of the initial distribution.  From this we obtain the $k$th cumulant (subscript $c$) as
\beq
  \ew{E^k(t)}_c &=& \left. \frac{\partial^k}{\partial(i\chi)^k} \mathcal{F}(\chi,t)\right|_{\chi=0}
  .
\eeq
Explicit expressions for the first two cumulants read 
\beq
  \ew{E(t)} &=& v_E t + E_0
  ;
  \nonumber\\
  \ew{E^2(t)}_c &=& 2 D_E t + \sigma_E^2
  \label{EQ:diffusion}
  ,
\eeq
where $E_0 = \ew{E(0)}$, $\sigma_E^2 = \ew{E^2(0)}_c$, and
\beq
  v_E &\equiv& - \int_0^\infty d \epsilon \, \epsilon \widetilde{\Gamma}(\epsilon)
  ;
  \nonumber\\
  D_E &\equiv&  \frac{1}{2} \int_0^\infty d \epsilon \, \epsilon^2 \widetilde{\Gamma}(\epsilon)
  \label{EQ:vEDEintegrals}
  .
\eeq
These last two quantities are the mean drift velocity of the electron energy distribution, and the diffusion constant \cite{Risken1996} associated with the spreading of the distribution.  
The higher cumulants of the $A^{(0)}$ distribution are non-zero (and could be calculated straightforwardly  within this approach). Indeed, a distribution starting as a Gaussian leaves a small tail at higher energies as it relaxes downwards. For relevant parameters, this tail is small and approximation of the complete behaviour as a Gaussian with just the first two cumulants provides a reasonably accurate description of the entire distribution.

\begin{figure}[t]
  \begin{center}
     \includegraphics[width=\columnwidth,clip=true]{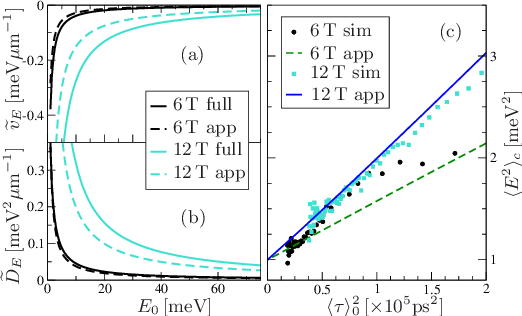}
  \end{center}
  \caption{  
    Drift-diffusion of the $A^{(0)}$ electron distribution.
    (a) the scaled energy-drift parameter $\widetilde{v}_E \equiv  v_E / v_0$ which gives the mean energy loss of the electron per distance travelled (in meV/$\mu$m). 
    (b) The similarly-scaled diffusion parameter $\widetilde{D}_E \equiv  2 D_E / v_0$ which gives how much the variance increases per distance travelled (in meV$^2$/$\mu$m). 
    Solid lines are  from the full integrals of \eq{EQ:vEDEintegrals}, whereas the dashed lines show the approximate forms from \eq{EQ:vEfinalTRI} and \eq{EQ:DEfinalTRI}.
    (c)  The variance $\ew{E^2}_c$ of the $A^{(0)}$ distribution as a function of mean time-of-flight-squared $\ew{\tau}_0^2$ for $B=6,12$ T.  The symbols represent results extracted from simulations; the straight lines are the analytic result from \eq{EQ:gradients} with coefficients calculated from the full integrals.
    \label{FIG:vE_and_DE}
  }
\end{figure}

The integrals of \eq{EQ:vEDEintegrals} can be computed numerically.  They can also be evaluated approximately using a saddle-point technique as outlined in Appendix\,\ref{SEC:APdiff}. This approximation yields the expressions
\beq 
  v_E 
  &\approx&
  -
   \frac{1}{16\sqrt{2\pi}}
  \frac{ c_{\mbox{\tiny LA}} M_{\mbox{\tiny LA}}^2}{a^3 l_\Omega \hbar v_0}
  ;
  \label{EQ:vEfinalTRI}
  \\
  D_E 
  &\approx&
  \frac{1}{4(2\pi)^{3/2}}
  \frac{c_{\mbox{\tiny LA}}^2 M_{\mbox{\tiny LA}}^2}{ a^4 l_\Omega v_0}
  \label{EQ:DEfinalTRI}
  .
\eeq
The derivation of these expressions also uses the fact that the ratio $c_{\mbox{\tiny LA}}/v_0$ is small.  Physically, this means that the phonons are emitted preferentially in a direction transverse to the electron propagation.
\fig{FIG:vE_and_DE} (a) and (b) show both full and approximate values for $v_E$ and $D_E$.  Here we have plotted $\widetilde{v}_E \equiv  v_E / v_0$ and $\widetilde{D}_E \equiv  2 D_E / v_0$, as these give the energy drop and variance-increase per unit distance travelled. We see that, except for close to the band bottom, the energy lost is a fraction of a meV per micron.  The diffusion constant gives a similar increase in variance per micron. These figures also show that our approximations capture the qualitative behaviour well across the energy range, but are quantitatively most accurate in the high-energy limit.

%%%%%%%%%%%%%%%%%%%%%%%%%%%%%%%%%%%%%%%%%%%%%%%%%%%%%%%%%%%%%%%%
%%%%%%%%%%%%%%%%%%%%%%%%%%%%%%%%%%%%%%%%%%%%%%%%%%%%%%%%%%%%%%%%

The only energy dependence in \eq{EQ:vEfinalTRI} and \eq{EQ:DEfinalTRI} is through the $1/v_0$ factor.  Thus, with time of flight $\tau= x_D/v_0$ we can rewrite Eqs.~(\ref{EQ:diffusion}) as
\beq 
  \ew{E(t)} &=& \lambda_1 \tau^2 + E_0;
  \nonumber\\
  \ew{E^2(t)}_c &=& \lambda_2 \tau^2 + \sigma_E^2
  \label{EQ:gradients}
  ,
\eeq
where 
\beq 
  \lambda_1  = \frac{v_E v_0}{x_D}
  ;\quad
  \lambda_2  = \frac{2 D_E v_0}{x_D}
  .
\eeq
are energy-independent gradients.  Thus plots of the energy drop and variance against the mean time of flight for the $A^{(0)}$ distribution should be straight lines.
\fig{FIG:vE_and_DE}(c) illustrates this for the energy variance.  To compare with simulations, we extract the $A^{(0)}$ distribution in the outer edge channel and renormalise with the survival probability $P_0$.  From this we extract the variance of the distribution and the mean time of flight.
In the high-energy range shown here (corresponding to low-$\tau^2$), the analytic theory matches the simulations well at high field (here $B=12$\,T).  Deviations from the analytic result for $B=6$\,T are apparent, but this is not surprising given that at lower fields we have significant scattering out of the $n=0$ edge channel.  We might expect further improvements in the agreement at even higher energies.  However, here the survival probability becomes very small and statistics from simulations becomes unreliable  (as would also be the case in experiment).  At lower energies (high $\tau^2$),  we come close to the band bottom, and the approximations used here cease to be valid.

%%%%%%%%%%%%%%%%%%%%%%%%%%%%%%%%%%%%%%%%%%%%%%%%%%%%%%%%%%%%%%%%%%%%%%%%%
%%%%%%%%%%%%%%%%%%%%%%%%%%%%%%%%%%%%%%%%%%%%%%%%%%%%%%%%%%%%%%%%%%%%%%%%%
%%%%%%%%%%%%%%%%%%%%%%%%%%%%%%%%%%%%%%%%%%%%%%%%%%%%%%%%%%%%%%%%%%%%%%%%%
\section{Discussion \label{SEC:discussions}}

From the point-of-view of probing fundamental semiconductor physics, single electron sources offer rich possibilities.   The results here show that the information encoded in the single-electron ATD can give significant insight into acoustic- and optical- phonon emission processes.
Nevertheless, for applications, and particularly applications in the arena of hot electron quantum optics, it would presumably be best if all inelastic processes could be ``switched off''.  
The results we have presented here show that there are significant experimental handles that can be used to modify and suppress both LO and LA interactions.  However, many of these handles effect the strength of the two interactions in opposite ways and thus care must be taken to find an optimal trade-off between the two.

At low injection energies, $E_0 < \hbar \omega_{\mbox{\tiny LO}}$, the lack of possible destination states means that LO-phonon emission is absent.  However, in this regime, the effects of LA-phonon emission are most pronounced, mainly because of the low velocity of electrons 
\footnote{
  The electron velocity is important on two accounts: faster electrons obviously have less time in the system during which to emit phonons; secondly, the phonon-emission rates depend inversely on the electron velocity thanks to the one-dimensional density of electronic states entering in \eq{EQ:ratedensity}.
}.
At higher energies, the relaxation effects of LA-phonons are considerably suppressed, mainly due the speed of transit of the electrons through the system.  However, higher energy means an increase in activity of LO-phonon emission.
The role of magnetic field strength also plays a conflicting role.  Whilst increasing $B$ suppresses the LO emission rate exponentially, it increases the effects of LA emission, and quantities such as mean energy drop and energy spread increase algebraically with magnetic field.  Thus it would appear that intermediate values of injection energy and magnetic field offer a good compromise.

More significantly, perhaps, is the dependence of the acoustic phonon emission rates on the strength of confinement in the $z$-direction, which enters our calculation through a  parameter $a$ that characterises the extent of wave function in this direction [see  \eq{EQ:FH}].  The energy lost to phonons scales like $\ew{\Delta E} \sim a^{-3}$, and the distribution width scales like $\sqrt{\ew{E^2}_c} \sim a^{-2}$ \footnote{A general energy cumulant scales like {$\ew{E^k}_c \sim 1/a^{k+2}$}}.
This means that the most prominent effects of acoustic phonon emission can be significantly reduced by using a wider well (and hence larger $a$) to define the two-dimensional electron gas. From \citer{Emary2016} we see that the LO-phonon rates are, to first approximation, independent of this width, and should thus be unaffected by this.

In the calculations presented here, we have only considered electron-acoustic-phonon scattering via the deformation potential interaction.  In GaAs, piezoelectric field scattering is also known to be significant in low-dimensional geometries in some parameter regimes \cite{Climente2006}.  We have repeated our simulations with these piezoelectric interactions present, and as we show in Appendix \ref{SEC:APPZ}, these generally have little influence. The only exception to this is at low fields in the region where significant population of the $n=1$ subband is observed at the detector. This population is increased slightly by the piezoelectric interactions. 

We have focussed here exclusively on the case where electrons are injected into the outermost edge channel, corresponding to the situation generally held to be the case in experiments with dynamic quantum dot sources \cite{Fletcher2013,Waldie2015,Johnson2017a}. 
Injection into the $n=1$ might one day be possible and we here comment on this possibility.  On the numerical side, it is clear that our approach can be extended to this scenario (e.g. \fig{FIG:LArates} contains the $n=1$ to $n'=1$ acoustic-phonon rate).
Within the analytic approach of Sec.~\ref{SEC:DD}, both drift and diffusion parameters in the $n=1$ subband are scaled by a factor of $1/\sqrt{3} $ relative to those within the $n=0$ subband.
However, the most significant effect of injecting into the $n=1$ subband is that the fast LO emission process $n=1 \to n'=0$ involved in the two-phonon emission described in Sec.~\ref{SEC:LO} is now a relaxation channel directly open to the hot electron, rather than needing first the emission of an LA phonon. For this reason we expect the LO emission rates from the inner channels to be significantly increased over those for the outermost  channel, see \citer{Emary2016} for more details.

The Monte-Carlo procedure that we have used to find the arrival-time distribution is manifestly a semi-classical one, where the rates are determined by quantum mechanics but the behaviour of the electron distribution is taken to be a fixed wavepacket around a classical trajectory.  As discussed in \citer{Emary2016} this approach is justified for time-of-flight type experiments with hot electrons, e.g. \cite{Johnson2018}, where the effects of quantum dispersion can largely be neglected. However, in future quantum-optics style experiments relying on interference this approach will need to be extended to treat coherences as well as populations. 
It also remains as future work to evaluate the extent to which residual interactions with the Fermi-sea electrons contribute to the relaxation and decoherence of these electrons.
In this context, we note that much progress has been made in understanding the decoherence and relaxation of low-energy single-electron sources through bosonization techniques, e.g. \citer{Degiovanni2009,Ferraro2014,Wahl2014}.

\begin{acknowledgments}
  We are grateful to H.-S.~Sim and A.~Dyson for useful discussions.  This research was supported by EPSRC grant EP/P034012/1.  M.~K. was supported by the UK department for Business, Energy and Industrial Strategy, and by the project EMPIR 17FUN04 SEQUOIA. This project has received funding from the EMPIR programme co-financed by the Participating States and from the European Union’s Horizon 2020 research and innovation programme.
\end{acknowledgments}

%%%%%%%%%%%%%%%%%%%%%%%%%%%%%%%%%%%%%%%%%%%%%%%%%%%%%%%%%%%%%%%%%%%%%%%%%
%%%%%%%%%%%%%%%%%%%%%      APPENDIX     %%%%%%%%%%%%%%%%%%%%%%%%%%%%%%%
%%%%%%%%%%%%%%%%%%%%%%%%%%%%%%%%%%%%%%%%%%%%%%%%%%%%%%%%%%%%%%%%%%%%%%%%%
\appendix

%%%%%%%%%%%%%%%%%%%%%%%%%%%%%%%%%%%%%%%%%%%%%%%%%%%%%%%%%%%%%%%%%%%%%%%%%%%%%%%%
%%%%%%%%%%%%%%%%%%%%%%%%%%%%%%%%%%%%%%%%%%%%%%%%%%%%%%%%%%%%%%%%%%%%%%%%%%%%%%%%
%%%%%%%%%%%%%%%%%%%%%%%%%%%%%%%%%%%%%%%%%%%%%%%%%%%%%%%%%%%%%%%%%%%%%%%%%%%%%%%%
\section{Electron-phonon matrix elements and confinement parameters \label{SEC:APmatrixeles}}
We take the matrix element for the interaction of electrons with LO phonons to be \cite{Mahan2000} 
\beq
  M_{\mbox{\tiny LO}}^2 = 4\pi \alpha \hbar \frac{\rb{\hbar \omega_\text{LO}}^{3/2}}{\rb{2m_e^*}^{1/2}}
  \label{EQ:M0alpha}
  ,
\eeq
with the following parameters for GaAs: electron effective mass, $m_e^*=0.067m_e$;
LO phonon energy, $\hbar \omega_\text{LO} = 36$\,meV; and coupling constant, $\alpha=0.068$.

For the interaction with acoustic phonons, we assume the LADP interactions with matrix element \cite{Climente2006}
\beq
  M_{\mbox{\tiny LA}}^2 = 
  \frac{\hbar\,D^2}{2\,d\,c_{\mbox{\tiny LA}}}
  ,
\eeq
with density $d=5310$ kgm$^{-3}$, crystal acoustic deformation potential $D=8.6$\,eV, and speed of sound $c_{\mbox{\tiny LA}}=4720$\,ms$^{-1}$.

In contrast to \citer{Emary2016}, which considered a square-well confinement in the $z$ direction, we consider here a triangular quantum well with ground-state wave function given by the Fang-Howard Ansatz  \cite{Ando1982}
\beq
  \phi_0(z) = \rb{2a^3}^{-1/2} z e^{-z/(2a)}
  \label{EQ:FH}
  ,
\eeq
where parameter $a$ determines the extent of the wave function.  Here we choose a value of $a=3$\,nm.  For the transverse harmonic confinement, we take a parameter of $\hbar\omega_y = 2.78$ meV, obtained as a fit to the $V_{G5}=-0.25$V time-of-flight results of \citer{Johnson2018}.

%%%%%%%%%%%%%%%%%%%%%%%%%%%%%%%%%%%%%%%%%%%%%%%%%%%%%%%%%%%%%%%%%%%%%%%%%%%%%%%%
%%%%%%%%%%%%%%%%%%%%%%%%%%%%%%%%%%%%%%%%%%%%%%%%%%%%%%%%%%%%%%%%%%%%%%%%%%%%%%%%
%%%%%%%%%%%%%%%%%%%%%%%%%%%%%%%%%%%%%%%%%%%%%%%%%%%%%%%%%%%%%%%%%%%%%%%%%%%%%%%%
\section{Evaluation of LADP emission rates\label{SEC:APrates}}

Analogous to the LO-phonon calculation of \citer{Emary2016}, the rates of \eq{EQ:FGR1} for LADP emission can be written
\beq
  \Gamma^{\mbox{\tiny LA}}_{n'k'nk}
  &=&
  \frac{2\pi}{\hbar}
  \sum_\mathbf{q}
  \left|\mathcal{M}_{\mbox{\tiny LA}}(\mathbf{q})\right|^2
  \left|G^{(z)}(q_z)\right|^2
  \left|G_{n'k',nk}^{(y)}(q_y)\right|^2
  \nonumber\\
  &&
  ~~~~~
  \times
  \delta_{q_x,k'-k}
  \delta\rb{
    E_{n'k'}-E_{nk} + \hbar c_{\mbox{\tiny LA}} q
  }
  \label{EQ:APFGR}
  ,
\eeq
with
\beq
  |G^{(y)}_{n'k',nk}(q_y)|^2
  =
  F^{(y)}_{n'n}
  \rb{\sqrt{
    \textstyle{\frac{1}{2}}l_\Omega^2\left[
     q_y^2 
      + \rb{\frac{\omega_c}{\Omega}}^2(k'-k)^2
    \right]
  }}
  \nonumber
  ,
\eeq
and
\beq
  F^{(y)}_{n'n}(Q) = \frac{n_<!}{n_>!}e^{-Q^2} Q^{2|n'-n|}\left[L_{n_<}^{|n'-n|}\rb{Q^2}\right]^2
  .
\eeq
We consider here a triangular quantum well in the $z$ direction, for which the structure factor reads 
\beq
 |G^{(z)}(q_z)|^2 = F^{(z)}(a q_z)
 ;
 ~~
  F^{(z)}(Q) = \frac{1}{(1+Q^2)^3}
  .
\eeq
To evaluate \eq{EQ:APFGR}, we take the continuum limit and switch to polar coordinates:
$q_x =  |q| \cos \theta$; $q_y = |q| \sin \theta \cos \phi$;  $q_z = |q| \sin \theta \sin \phi$.
Writing
\beq
  q_0 = \rb{E_{nk} - E_{n'k'}}/{\hbar c_{\mbox{\tiny LA}}}
  ,
\eeq
and 
\beq
  \cos\theta_0 = \rb{k'-k}/{q_0};
  \qquad \theta_0 \in [\pi/2,\pi],
\eeq
we reduce \eq{EQ:APFGR} to 
\beq
  \Gamma^{\mbox{\tiny LA}}_{n'k'nk}
  &=&
  \frac{1}{2\pi  \hbar^2 c_{\mbox{\tiny LA}} L}
  \int_0^{2\pi} d\phi \,
  q_0
  \left|L^{3/2}\mathcal{M}_{\mbox{\tiny LA}}(q_0,\theta_0,\phi)\right|^2  
  \nonumber\\
  &&
  \times
  F^{(z)}(a q_0 \sin \theta_0 \sin \phi)
  \left|G_{n'k'nk}^{(y)}(q_0 \sin \theta_0 \cos \phi)\right|^2 
  \nonumber\\
  &&
  \times\Theta\rb{E_{nk} -E_{n'k'} - \hbar c_{\mbox{\tiny LA}} |k'-k|}
  \nonumber\\
  &&
  \times\Theta\rb{E_{n'k'} - n' \hbar\Omega}
  \nonumber
  .
\eeq
The first unit-step function $\Theta$ here arises from conservation of energy and momentum [see \eq{EQ:constraint}], and the second avoids $E'$ coming in lower than the band bottom.

%%%%%%%%%%%%%%%%%%%%%%%%%%%%%%%%%%%%%%%%%%%%%%%%%%%%%%%%%%%%%%%%%%%%%%%%%%%%%%
%%%%%%%%%%%%%%%%%%%%%%%%%%%%%%%%%%%%%%%%%%%%%%%%%%%%%%%%%%%%%%%%%%%%%%%%%%%%%%
%%%%%%%%%%%%%%%%%%%%%%%%%%%%%%%%%%%%%%%%%%%%%%%%%%%%%%%%%%%%%%%%%%%%%%%%%%%%%%
\section{Estimation of diffusion parameters \label{SEC:APdiff}}
The linearisation of \eq{EQ:linearise} means that for $n=n'=0$ we have 
$
 \cos \theta_0 = - {c_{\mbox{\tiny LA}}}/{v_0}
$,
which is constant in this approximation, and also that the $(k-k')^2$ in the argument of $G_{n'k'nk}^{(y)}$ becomes $(E-E')^2/(\hbar v_0)^2$.  Thus we see that the expression for the rate only depends on the difference $\epsilon=E-E'$.

%%%%%%%%%%%%%%%%%%%%%%%%%%%%%%%%%%%%%%%%%%%%%%%%%%%%%%%%%%%%%%%%
\begin{figure}[tb]
  \begin{center}
     \includegraphics[width=\columnwidth,clip=true]{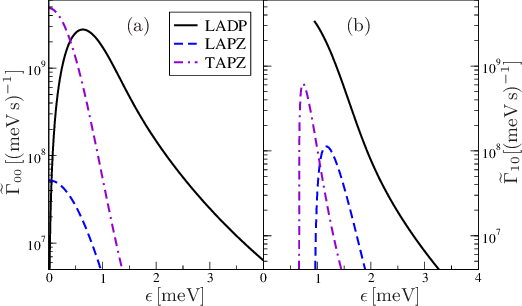}
  \end{center}
  \caption{ 
    Phonon emission rate densities $\widetilde{\Gamma}_{n'n}(E-\epsilon,E)$ as a function of energy loss $\epsilon$ for the three  acoustic phonon interactions:  LADP (solid black lines), LAPZ (blue dashed), and TAPZ (violet dashed-dot).  Left: transitions within the outermost LL $n=0\to n'=0$. Right: first inward transition $n=0\to n'=1$. The magnetic field was $B=12$\,T and the starting energy was $E = 100$\,meV.
    \label{FIG:APrates}
  }
\end{figure}
%%%%%%%%%%%%%%%%%%%%%%%%%%%%%%%%%%%%%%%%%%%%%%%%%%%%%%%%%%%%%%%%

Introducing the energy scale
$
  E_z = {\hbar c_{\mbox{\tiny LA}}}/{a}
$, the diffusion parameter in this approximation becomes
\beq    
 D_E &\approx&  
 \frac{
    M_{\mbox{\tiny LA}}^2
  }{
    4 \pi \hbar^3 a^2 c_{\mbox{\tiny LA}} v_0
  }
  E_z^3 I_D
  ,
\eeq
with the integral
\beq
  I_D
  &\equiv&
  \frac{1}{\pi}
  \int_0^\infty d \tilde{\epsilon} \,  
  \int_{-\pi/2}^{\pi/2} d\phi \,\,
  \tilde{\epsilon}^4  \exp\left[
    - \textstyle{\frac{1}{2}}r_V^2 \tilde{\epsilon}^2  
  \right]
  \nonumber\\
  &&\times
  F^{(z)}\rb{\tilde{\epsilon}\sin \theta_0\cos\phi} 
%   \nonumber\\
%   &&\times
   \exp\left[
    -\textstyle{\frac{1}{2}} r_A^2 \tilde{\epsilon}^2\sin \theta_0 \sin^2\phi
  \right] 
  \nonumber
  .
\eeq
Here
$
  \tilde{\epsilon} \equiv (E-E')/E_z
$,
is the dimensionless energy loss, and we have defined
and the ratios 
\beq
  r_V = \frac{\omega_c c_{\mbox{\tiny LA}}l_\Omega}{\Omega v_0a}
  ;\quad \mathrm{and} \quad
  r_A  = \frac{l_\Omega}{a}
  .
\eeq
To obtain an approximate analytic form for this integral we first note that, for typical parameters, $r_V\sim 1/50$ compared with $r_A \sim 1/2$.  We thus drop the initial exponential factor. We then approximate the integral over $\phi$ with a saddle-point approximation about $\phi=\pi/2$.  This gives 
\beq
  I_D
  &\approx& 
  \sqrt{\frac{2}{\pi}}\frac{1}{r_A}\frac{1}{\sin^5\theta_0}
  \int_0^\infty d u \, 
  u^3
  F^{(z)}\rb{u} 
  \label{EQ:ID}
  .
\eeq
For the triangular well, this last integral can be performed analytically and we obtain
\beq 
  I_D
  \approx
  \frac{1}{2 \sqrt{2\pi}r_A \sin^5\theta_0}
  .
\eeq
Finally, since for typical injection energies $c_{\mbox{\tiny LA}}/v_0 \ll 1$, we can approximate
$
   \sin \theta_0 \approx 1 + \mathcal{O}\rb{c_{\mbox{\tiny LA}}^2/v_0^2},
$
such that we obtain \eq{EQ:DEfinalTRI}.  A similar calculation for the energy drift-velocity yields
\eq{EQ:vEfinalTRI}.

%%%%%%%%%%%%%%%%%%%%%%%%%%%%%%%%%%%%%%%%%%%%%%%%%%%%%%%%%%%%%%%%%%%%%%%%%%%%%%%%
%%%%%%%%%%%%%%%%%%%%%%%%%%%%%%%%%%%%%%%%%%%%%%%%%%%%%%%%%%%%%%%%%%%%%%%%%%%%%%%%
%%%%%%%%%%%%%%%%%%%%%%%%%%%%%%%%%%%%%%%%%%%%%%%%%%%%%%%%%%%%%%%%%%%%%%%%%%%%%%%%
\section{LAPZ and TAPZ interactions \label{SEC:APPZ}}

The two additional acoustic-phonon scattering mechanisms in GaAs are electron scattering due to the piezoelectric field by longitudinal phonons (LAPZ interaction) and transverse phonons (TAPZ interaction).  These can be incorporated in our calculations with additional interaction terms in the Hamiltonian similar in structure to the acoustic-phonon terms in \eq{EQ:FH1} with matrix elements \cite{Climente2006}:
\beq
  \lvert M_{\mbox{\tiny LAPZ}}(\mathbf{q}) \rvert^2=\frac{32 \pi^2\,\hbar\,e^2\,h_{14}^2}{\varepsilon_r^2\,d\,c_{\mbox{\tiny LA}}\,L^3}\,
  \frac{(3\,q_x\,q_y\,q_z)^2}{q^7}
  ,
\eeq
and
\beq
  \lvert M_{\mbox{\tiny TAPZ}}(\mathbf{q}) \rvert^2
  &=&\frac{32 \pi^2\,\hbar\,e^2\,h_{14}^2}{\varepsilon_r^2\,d\,c_{\mbox{\tiny TA}}\,L^3}
  \nonumber\\
  &&
%   \!\!\!\!\!\!\!\!\!\!
  \times
  \left| \frac{q_x^2\,q_y^2 + q_y^2\,q_z^2 + q_z^2\,q_x^2}{q^5} - \frac{(3\,q_x\,q_y\,q_z)^2}{q^7} \right|
  .
  \nonumber
\eeq
For parameter values we take $c_{\mbox{\tiny TA}}=3340$ ms$^{-1}$, 
$\varepsilon_r=12.9$, and $h_{14}=1.41\times 10^9$ Vm$^{-1}$.

%%%%%%%%%%%%%%%%%%%%%%%%%%%%%%%%%%%%%%%%%%%%%%%%%%%%%%%%%%%%%%%%
\begin{figure}[tb]
  \begin{center}
     \includegraphics[width=\columnwidth,clip=true]{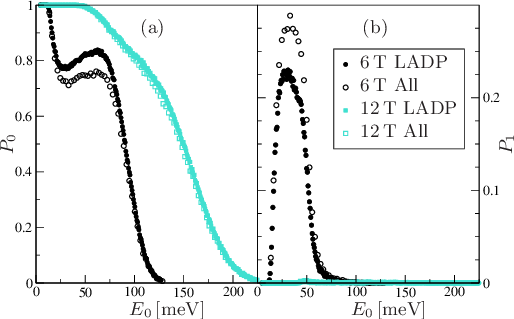}
  \end{center}
  \caption{
    Survival probabilities (a) $P_0$ and (b) $P_1$ as a function of injection energy $E_0$ for two magnetic field strengths, $B=6,12$ T.  
    Solid symbols show results with LO + LADP interactions only. Open symbols show results with LO + all three acoustic phonon interactions.
    \label{FIG:APSP3phonon}
  }
\end{figure}
%%%%%%%%%%%%%%%%%%%%%%%%%%%%%%%%%%%%%%%%%%%%%%%%%%%%%%%%%%%%%%%%

\fig{FIG:APrates} shows the rate densities for these interactions in comparison with those for the LADP interaction for $B=12$\,T.  For the inter-channel scattering (determinant for the LA+LO knee in the effective rate), the LAPZ and TAPZ rate are clearly much smaller than the LADP rate.  
For the intra-channel case, the situation is a bit more complicated.  The LAPZ interaction is clearly negligible compared with LADP.  For most of the range of energy-loss $\epsilon$, the same is true for the TAPZ interaction. However, for $\epsilon \to 0$, the TAPZ rate density tends to a constant value, whereas the LADP density falls to zero.  That this excess of the TAPZ rate density at small energy changes is generally unimportant can be appreciated by considering the cumulant integrals such as in \eq{EQ:vEDEintegrals}.  Since the integrand for the $k$th cumulant involves a factor $\epsilon^{k+1}$ [see \eq{EQ:ID}], the magnitude of the rate at $\epsilon \to 0$ does not contribute to any cumulant, and hence the behaviour of the entire energy distribution.
The influence of the two piezoelectric interactions on the results presented here can be appraised from \fig{FIG:APSP3phonon}, which shows the survival probabilities $P_{0,1}$ calculated both with and without them.  For $B=12$\,T, the piezoelectric terms make no appreciable difference.  

At lower field ($B=6$\,T) the strength of the two piezoelectric interactions relative to the LADP interaction increases.  The overall effect of this is again small, except for the region where we have significant population of the $m=1$ level.  Here the piezoelectric processes lead to a small increase in this population.

%%%%%%%%%%%%%%%%%%%%%%%%%%%%%%%%%%%%%%%%%%%%%%%%%%%%%%%%%%%%%%%%%%%%%%%%%
%%%%%%%%%%%%%%%%%%%%%      REFERENCES     %%%%%%%%%%%%%%%%%%%%%%%%%%%%%%%
%%%%%%%%%%%%%%%%%%%%%%%%%%%%%%%%%%%%%%%%%%%%%%%%%%%%%%%%%%%%%%%%%%%%%%%%%
\bibliography{/home/gloygum/Dropbox/Bibliography/CEbib.bib}
\end{document}